\documentclass[article,pt11,twocolumn]{revtex4-1}
\usepackage{graphicx}
\usepackage{dcolumn}
\usepackage{amsmath}
\usepackage{amsfonts}
\usepackage{upgreek}
\usepackage{color}
\usepackage{natbib}
 \usepackage{float}
\begin{document}

\title{\textsf{Skyrmion Lattice Collapse and Defect-Induced Melting in Chiral Magnetic Films}}

\author{L. Pierobon\textsuperscript{1}, C. Moutafis\textsuperscript{2}, Y. Li\textsuperscript{2}, J. F. L\"offler\textsuperscript{1}, M. Charilaou\textsuperscript{1}}
\affiliation{\textsuperscript{1} Laboratory of Metal Physics and Technology, Department of Materials, ETH Zurich, Zurich 8093, Switzerland \linebreak
\textsuperscript{2} School of Computer Science, University of Manchester, M13 9PL Manchester, UK}

\begin{abstract}

Magnetic phase transitions are a test bed for exploring the physics of non-equilibrium phenomena in condensed matter, which become even more complex when topological constraints are involved. In particular, the investigation of skyrmions and skyrmion lattices offers insight into fundamental processes of topological-charge creation and annihilation upon changing the magnetic state. Nonetheless, the exact physical mechanisms behind these phase transitions remain unresolved. Here, we have systematically compared ultra-thin films with isotropic and anisotropic Dzyaloshinskii-Moriya interactions (DMI), demonstrating a nearly identical behavior in technologically relevant materials such as interfacial systems. We numerically show that in perfect systems skyrmion lattices can be inverted in a field-induced first-order phase transition. The existence of even a single defect, however, replaces the inversion with a second-order phase transition of defect-induced lattice melting. This radical change in the system's behavior from a first-order to a second-order phase transition signifies the importance of such an analysis for all realistic systems in order to correctly interpret experimental data. Our results shed light on complex topological charge annihilation mechanisms that mediate transitions between magnetic states and pave the way for an experimental realization of these phenomena.
\end{abstract}

\maketitle
\section{Introduction}

The interplay between the symmetric Heisenberg exchange and antisymmetric Dzyaloshinskii-Moriya interaction \cite{Moriya1960} (DMI), and long-range magnetostatic interactions generates complex spin textures, such as helical, conical, and as shown recently \cite{Roesler2006,Bode2007,nagao2013} skyrmion phases. Magnetic skyrmions are particle-like spin configurations that occur on surfaces and interfaces upon rotational-symmetry breaking by either an external magnetic field \cite{Buhrandt2013} or perpendicular magnetic anisotropy (PMA) \cite{Wilson2014}. The skyrmion size depends on the strength of these fields, and their handedness on the type and sign of DMI \cite{Ezawa2010,Guslienko2015,Rohart2013}, but regardless of these properties they always have a topological charge $Q$ equal to 

\begin{equation} 
\label{eq:topcharge} 
Q=\frac{1}{4 \pi} \int {\bf m} \cdot ( \partial _{x} {\bf m} \times \partial _{y} {\bf m} ) dx dy = \pm 1\; ,
\end{equation} 

\noindent with $\bf{m}$ the magnetization unit vector. This signifies that for both Bloch- and N\'eel-type skyrmions \cite{hellman2017} the local magnetic moment rotates by $2\pi$ from one end of a skyrmion to another, as described by a variational ansatz \cite{Braun2012} for a 2$\pi$ domain wall.

On a surface containing more than one skyrmion, the global topological charge is the net sum resulting from all topological objects in the system. Skyrmions can be arranged in irregular clusters  \cite{Li2016} or rectangular lattices \cite{Heinze2011}, but mostly hexagonal lattices, as observed for the chiral magnets MnSi \cite{Muhlbauer2009,Tonomura2012}, FeCoSi \cite{Yu2010} and FeGe \cite{Yu2010a}. Skyrmion lattices (SkL) are stable close to the Curie temperature in bulk systems with surfaces, and in a much wider temperature range in thin films \cite{Yu2010a,Huang2012}. Additionally, in ultra-thin systems with interface-induced anisotropic DMI, such as Ir/Co/Pt multilayers \cite{Moreau-Luchaire2016,Boulle2016}, skyrmions form even at room temperature and zero external magnetic field due to strong PMA. This, together with the fact that they can be controlled by relatively small current-densities \cite{jonietz2010}, makes skyrmions promising candidates for future spin-based applications \cite{Parkin2008,Sampaio2013,Fert2013,Fert2017}.

Apart from being technologically relevant, skyrmionic spin textures offer the potential for detailed investigations into topological aspects of magnetism \cite{nakajima2017}. Specifically, since skyrmion textures are protected by their topological constraints, they cannot be continuously unwound into a trivial ferromagnetic (FM) configuration athermally without a phase transition. The exact mechanism of this kind of phase transitions, however, remains relatively unexplored so far.

In the following we demonstrate using micromagnetic simulations that skyrmion lattices undergo a magnetic field-induced phase transition where an antiskyrmion is created for each skyrmion, which results in a transient $Q=0$ state and enables the switching of the lattice polarity. However, in the presence of even a single defect, this phase transition is replaced by a melting-type mechanism, where topological charge is gradually lost.

\section{Computational modeling}

We have performed high-resolution micromagnetic simulations to investigate the phase transition of skyrmion lattices in detail. In our simulations, we have studied ultrathin films where the skyrmion lattice phase is stable at temperatures low enough for our micromagnetic simulations to be valid, as we did not consider finite-temperature effects.

The total energy density of the system consists of: (i) ferromagnetic exchange; (ii) perpendicular magnetic anisotropy (PMA); (iii) isotropic (bulk) or anisotropic (interfacial) Dzyaloshinskii-Moriya interaction (DMI); (iv) Zeeman coupling to an external magnetic field; and (v) dipole-dipole interactions: 

\begin{align} \label{eq:energy} 
F=& A ({\bf \nabla \cdot m})^2+ D {\bf m }\cdot ({\bf \nabla \times m})-K_\textrm{u}(m_z)^2 \nonumber \\
&- \mu_0 M_\textrm{s}{\bf m \cdot H}_{\textrm{ext}}-\frac{\mu_0 M_\mathrm{s}}{2}{\bf m \cdot H}_{\textrm{demag}} \; ,
\end{align} 

\noindent where ${\bf m}={\bf M}/M_\textrm{s}$ is the magnetization unit vector with $M_\textrm{s}$ the saturation magnetization, $A$ is the exchange stiffness, $D$ is the strength of DMI (either bulk or interfacial), $K_\textrm{u}$ is the first-order uniaxial anisotropy constant, $\bf {H}_{\textrm{ext}}$ is the external magnetic field, and $\bf{H}_{\textrm{demag}}$ is the local demagnetizing field due to dipole-dipole interactions. The $z$-component of the magnetization is perpendicular to the film plane.

We compute the magnetic state by solving the Landau-Lifshitz-Gilbert (LLG) equation of motion 

\begin{equation} \label{eq:LLG} 
\partial _{t} {\bf m}=-\gamma ({\bf m \times H}_{\textrm{eff}})+\alpha({\bf m \times} \partial _{t} {\bf m}) \; ,
\end{equation} 

\noindent where $\gamma$ is the electron gyromagnetic ratio, $\alpha$ is the dimensionless damping parameter, and ${\bf H}_{\textrm{eff}}=-{\bf \partial _{m}} F / \mu _{0} M_\textrm{s}$ is the effective magnetic field in the material consisting of external and internal magnetic fields, which depend on the material parameters. The LLG simulations have been done with mumax3 \cite{Vansteenkiste2014}, a finite-difference GPU-based program.

For both isotropic and anisotropic DMI systems we have considered a wide range of material parameters within which values for most materials fall \cite{Beg2015,Yang2016,Moreau-Luchaire2016,Ericsson1981,YAMADA2003}: $D=1.0 - 2.0$ mJ/m\textsuperscript{2} and $K_\textrm{u}=20-800$ kJ/m\textsuperscript{3}. 

The thin films are discretized in a 480 $\times$ 480 $\times$ 2 mesh (sample dimensions 832 nm $\times$ 960 nm $\times$ 4 nm) with periodic boundary conditions in \textit{x}- and \textit{y}-directions. Additionally, different cell sizes (always less than half the exchange length \cite{Braun2012} $\delta_\mathrm{ex}=\sqrt{2 A_\mathrm{ex}/\mu_0 M_\mathrm{s}^2}\approx 10$~nm) were used to verify the numerical stability of the simulations. 

The external magnetic field is always applied and swept perpendicular to the film plane, and the quantities recorded at each field step are: ${\bf m}$, ${\bf M}$, $Q$, the total energy density of the system and the individual contributions to it.

\section{Results}

We start by analyzing the magnetization profile of a Bloch and a N\'eel skyrmion in zero magnetic field in thin films with DMI and PMA. As shown in Figure 1a-c, the \textit{z}-component of the magnetization for the two kinds of skyrmions is identical and can be described precisely by the variational ansatz for a 2$\pi$ domain wall \cite{Braun2012} corresponding to a topological charge of unity, as also calculated from the relaxed spin texture. In our system, skyrmions are configured in a hexagonal lattice (see Figure 1d-e) and the global topological charge is equal to the number of skyrmions in the lattice, which in our case is $Q=64$.

\begin{figure}[h]
\centering
\includegraphics[width=1\columnwidth]{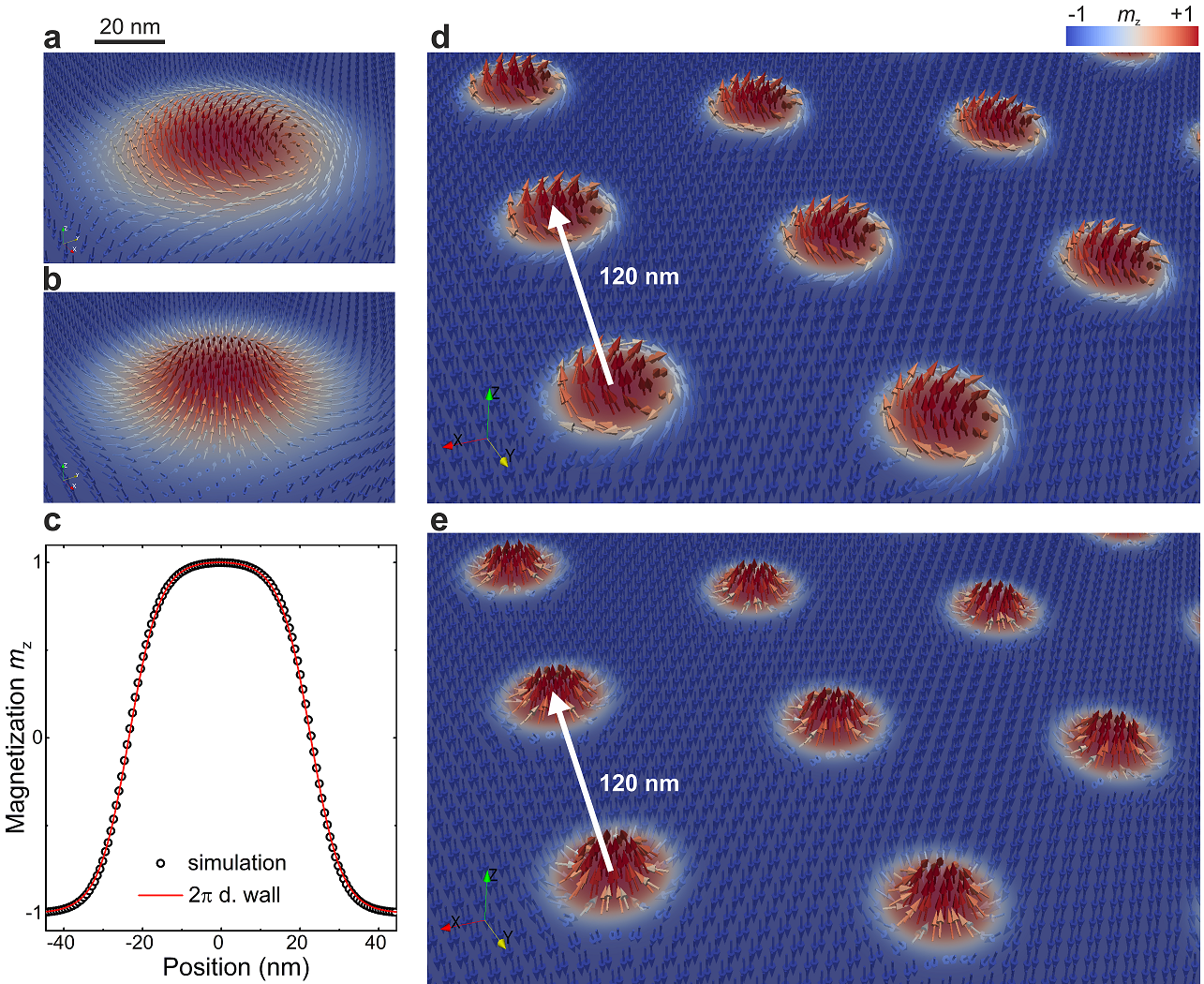}
\caption{Vector plots of \textbf{a} Bloch and \textbf{b} N\'eel skyrmions, which are found in materials with isotropic and anisotropic DMI respectively. \textbf{c} The magnetization \textit{z}-component profiles for the two kinds of skyrmions are identical and show an excellent agreement with the magnetization profile of a $2\pi$ domain wall (solid line). Skyrmions form hexagonal lattices with 6-fold symmetry, as shown in close-up vector plots of \textbf{d} Bloch and \textbf{e} N\'eel skyrmion lattices, where the white arrow marks the distance between neighboring skyrmions.}
\label{magnetization_profile}
\end{figure}

In order to investigate changes in skyrmion lattices as a function of external stimuli, we applied an external out-of-plane magnetic field by performing a magnetic field sweep in the direction both parallel and antiparallel to the skyrmion-core polarization, which we define as the positive and negative directions respectively. We observe that skyrmions shrink and consequently the net magnetization along the \textit{z}-axis decreases as the field is swept antiparallel to the core polarization (Figure 2a-d), as expected\cite{Moreau-Luchaire2016}. In a critical field $-|\textbf{\textrm{H}}_{\textrm{AP}}|$, the skyrmions are annihilated and the system reaches the topologically trivial FM state through a first-order phase transition. If the field is swept parallel to the core polarization, however, the resulting behavior is much more complex. Upon increasing the field parallel to the skyrmion-core polarization, skyrmions grow to the point where their boundaries themselves become a 2$\pi$ domain wall, meaning the topological charge is delocalized on a network of these domain walls. At this point skyrmions cannot grow anymore so they form a hexagonal state where the 2$\pi$ boundaries assume the lattice symmetry, as shown in Figure 2e. 

\begin{figure*}[!ht]
\centering
\includegraphics[width=1.8\columnwidth]{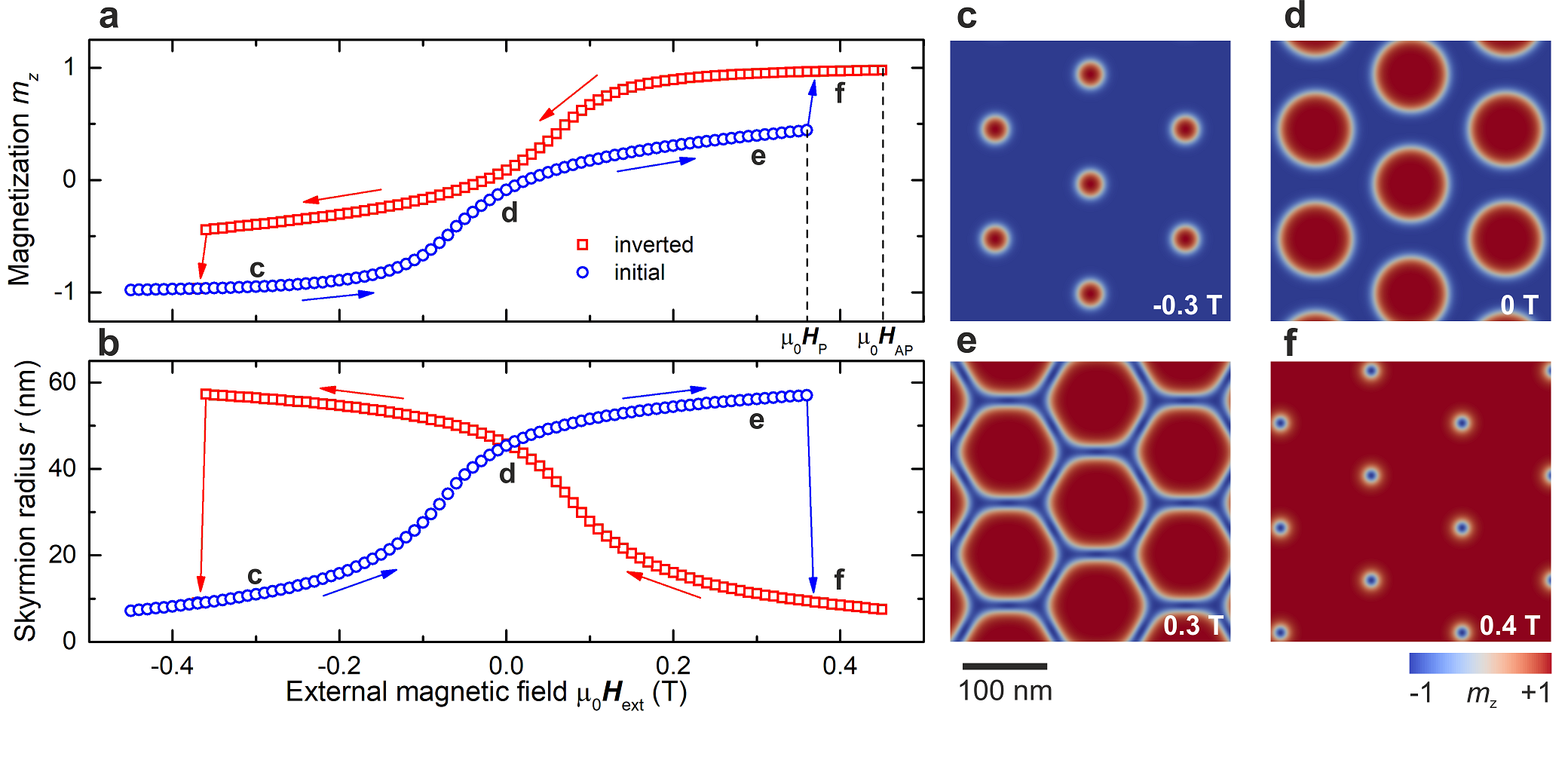}
\caption{Plots of \textbf{a} $m_{\textrm{z}}$ and \textbf{b} the radius of skyrmions as the magnetic field is swept. The inversion happens at $\pm|\textit{\textbf{H}}_{\textrm{P}}|$, and the collapse to the FM state happens at $\pm|\textit{\textbf{H}}_{\textrm{AP}}|$. Contour plots of the skyrmion lattice magnetization show \textbf{c-d} the growth of skyrmions in the parallel sweep with the topological charge localized in the centers of skyrmions, \textbf{e} the hexagonal state with the topological charge delocalized on the 2$\pi$ domain-wall boundary and \textbf{f} the inverted skyrmion lattice. The external magnetic field is indicated in the bottom right-hand corner of the plots and the positive direction of the field is defined as that of the skyrmion-core polarization in zero field. }
\label{hysteresis_loops}
\end{figure*}

From this state onwards there are two possible scenarios at a critical field $+|\textbf{H}_{\textrm{P}}|$: i) the system undergoes a first-order phase transition from the hexagonal to the FM state; or ii) the system undergoes another very surprising first-order phase transition in which the skyrmion lattice inverts its magnetic polarity (Figure 2f). This abrupt metamagnetic-like transition is characterized by a discontinuous change in the total magnetization and skyrmion radius (Figure 2a-b). As the field is swept further, skyrmions shrink and the inverted lattice undergoes a first-order phase transition to the FM state at $+|\textbf{H}_{\textrm{AP}}|$, in analogy to the destruction of the lattice in the antiparallel sweep. 

\begin{figure}[h]
\centering
\includegraphics[width=1.0\columnwidth]{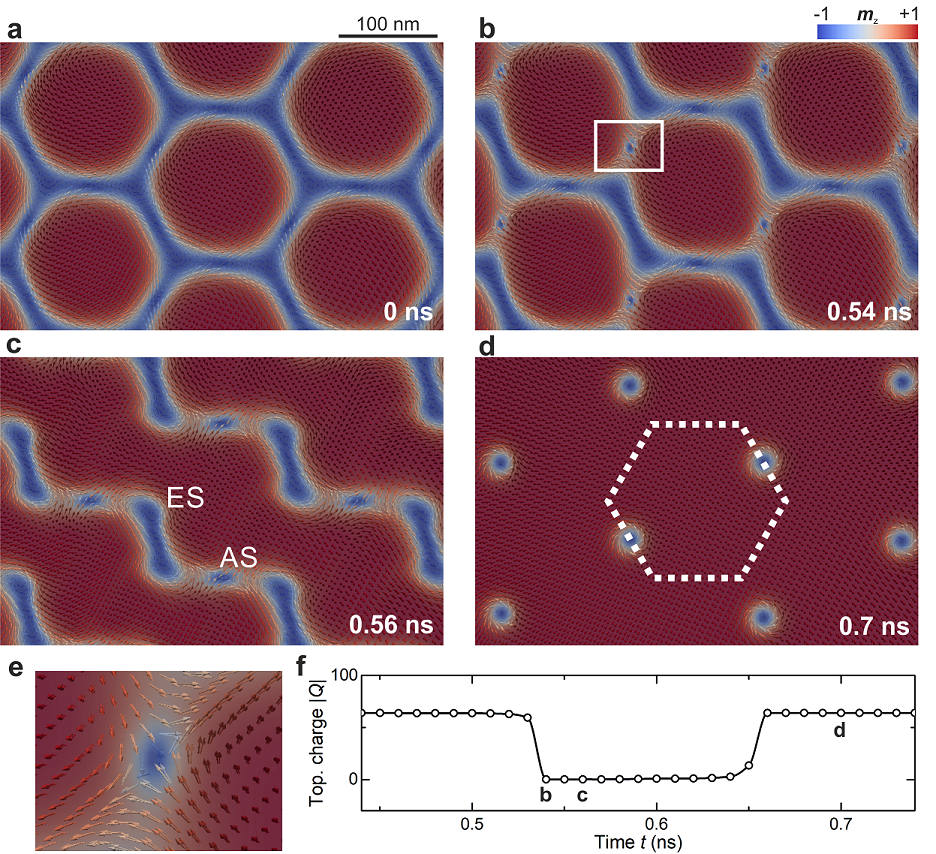}
\caption{Vector plots of $m_{\textrm{z}}$ showing the time-progression of the inversion at a field slightly higher than $+|\textit{\textbf{H}}_{\textrm{P}}|$ (time indicated in the bottom right-hand corner). \textbf{a} The hexagonal state collapses into \textbf{b-c} a transient state, which contains antiskyrmions and elliptical skyrmions (marked with AS and ES respectively in \textbf{c}). Antiskyrmions are annihilated at \textbf{d} the end of inversion and the white dotted hexagon shows the position of a skyrmion before the inversion to indicate the shift of the inverted lattice w.r.t. the original lattice. \textbf{e} A close-up of an antiskyrmion, marked in the white rectangle in \textbf{b}. \textbf{f} The temporal evolution of the global topological charge $Q$ shows that it vanishes in the transient state due to negatively charged antiskyrmions.}
\label{inversion}
\end{figure}

The metamagnetic transition of skyrmion-lattice inversion entails striking features related to the topological charge. As illustrated in Figure 3a-b, the inversion starts by a breaking of one third of the boundaries between skyrmions. Their 2$\pi$-domain-wall shape has a topologically constraining character, which induces a creation of an antiskyrmion for each skyrmion, exactly offsetting the global topological charge. In the next step, these antiskyrmions are annihilated and new pairs of elliptical skyrmions and antiskyrmions are created from the remaining boundaries (Figure 3c), keeping the global topological charge at zero. At the end of inversion, all antiskyrmions become annihilated, which restores the global topological charge to its original value. New cores of inverted skyrmions form from the elliptical skyrmions, which are situated at the vertices of skyrmions in the original SkL (Figure 3d), i.e., the inverted skyrmions have emerged from the 2$\pi$ boundaries. 

\begin{figure}[h]
\centering
\includegraphics[width=0.9\columnwidth]{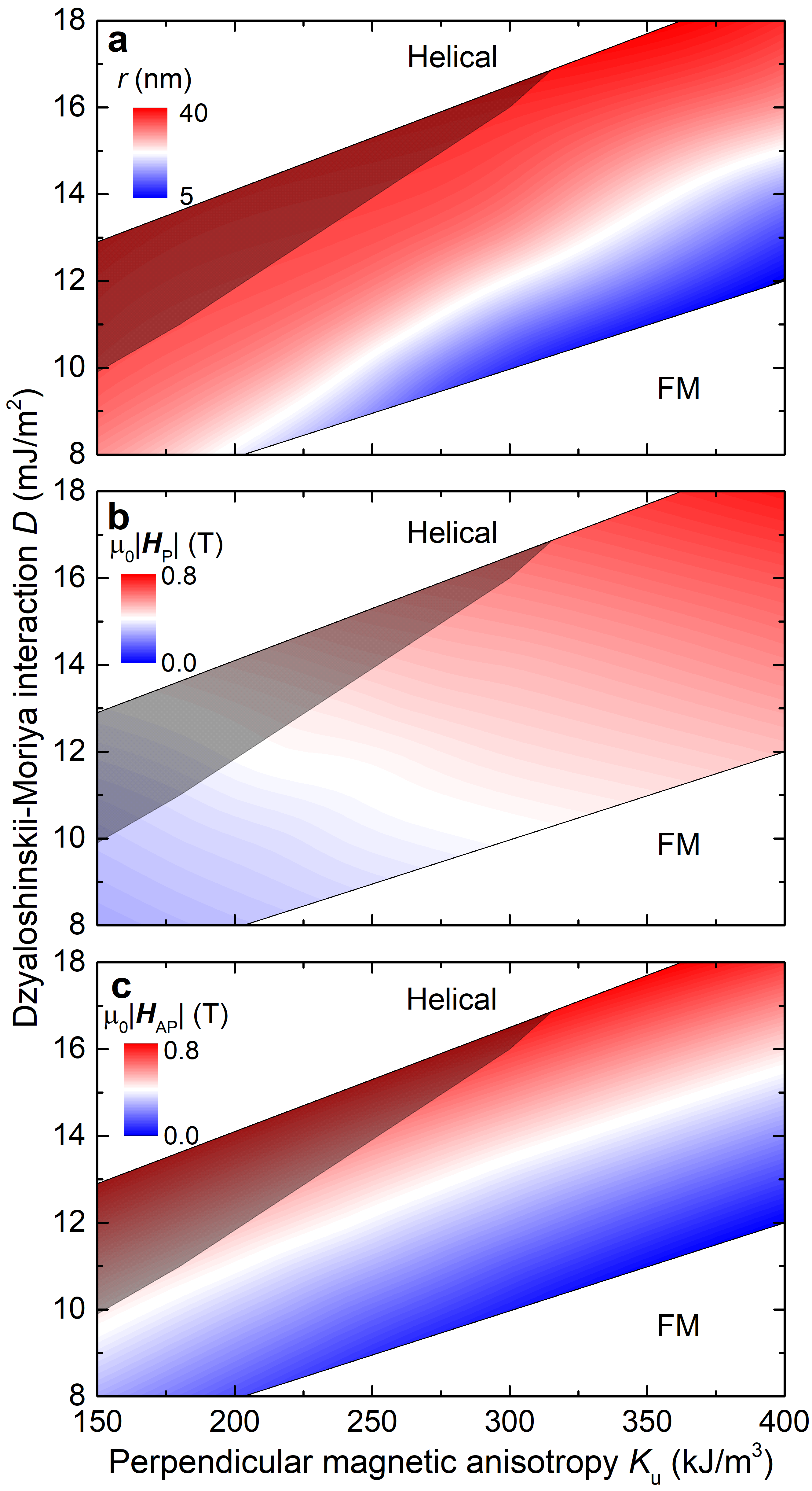}
\caption{Contour plots showing the range of PMA and DMI in which skyrmion lattices (colored) are stable in zero-magnetic field. The helical phase in the diagram is stable above the skyrmion lattice phase, and the ferromagnetic phase below. \textbf{a} Skyrmions are larger in materials with strong DMI and low PMA. \textbf{b} $|\textit{\textbf{H}}_{\textrm{P}}|$ increases with both PMA and DMI, and \textbf{c} $|\textit{\textbf{H}}_{\textrm{AP}}|$ increases with DMI but decreases with PMA. This means that the inversion happens only in the materials with low PMA (represented with the shaded area in the diagrams).}
\label{DMI_vs_Ku}
\end{figure}

The phenomena described above depend strongly on the intrinsic material parameters. For DMI strength between 1 and 2 mJ/m$^2$, comparable to the material values of intrinsically chiral magnets or Co/Pt-based multilayers, we have investigated the range of PMA i) for which skyrmion lattices are stable in zero magnetic field; and ii) for which the inversion happens. We compared bulk and interfacial systems, i.e., systems with isotropic and anisotropic DMI respectively, and found a nearly identical behavior. The SkL phase is stable in the range of $K_\mathrm{u}=150$ -- $800$ kJ/m$^3$ and the skyrmions grow with decreasing PMA and increasing DMI (Figure 4a), in agreement with literature \cite{Ezawa2010,Guslienko2015,Rohart2013}. The inversion can only occur if the inverted SkL is stable in the field range $|\textbf{H}_{\textrm{P}}|<\textbf{H}<|\textbf{H}_{\textrm{AP}}|$. The values of $|\textbf{H}_{\textrm{P}}|$ and $|\textbf{H}_{\textrm{AP}}|$, and therefore whether or not the SkL inversion occurs, depend on the material properties. As shown in Fig 4b-c, $|\textbf{H}_{\textrm{P}}|$ increases with increasing PMA, but $|\textbf{H}_{\textrm{AP}}|$ decreases with increasing PMA, so that the criterion limits the occurrence of inversion to materials with low PMA (shaded area in Figure 4).

\begin{figure*}[!ht]
\centering
\includegraphics[width=1.4\columnwidth]{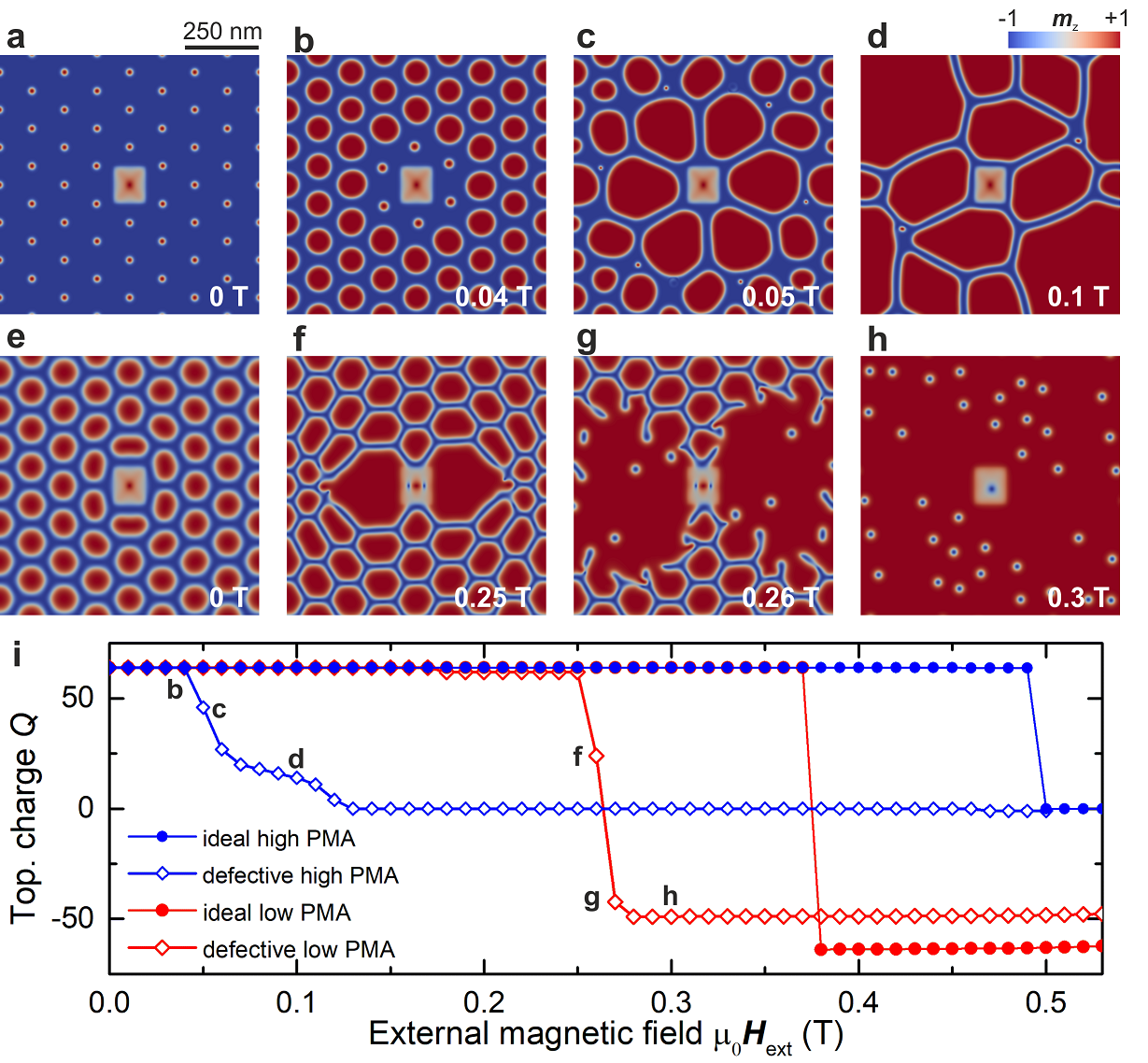}
\caption{Contour plots of $m_{\textrm{z}}$ showing defect-induced melting due to a local distortion of PMA in the region of a single skyrmion in a material with \textbf{a-d} PMA = 350 kJ/m\textsuperscript{3} (no inversion in the ideal case) and \textbf{e-h} PMA = 180 kJ/m\textsuperscript{3} (exhibits inversion in the ideal case). The value of the external magnetic field in each plot is marked in the bottom right-hand corner. The comparison between the ideal and defective lattices is shown in \textbf{i} the plot of $Q$ against the external field. Note that defect-induced melting does not happen if the magnetic field is swept antiparallel to the skyrmion core polarization; instead, skyrmions shrink until the FM state is reached just like in the perfect system.}
\label{defected_vs_perfect}
\end{figure*}

The findings described above assume an infinite ideal system. However, real materials, either bulk or thin-films, contain defects that could strongly affect the magnetic state, particularly in multilayers with interfacial DMI. It is therefore important to study the effect of defects on the SkL phase and its stability, which is essential for enabling the functionality of skyrmion-based devices \cite{Parkin2008,Sampaio2013,Fert2013,Fert2017}.

In our simulations, we have implemented defects in three different ways: i) a local variation of DMI or PMA, ii) a local distortion in the skyrmion lattice, and iii) a vacancy in the skyrmion lattice. We find that while defects do not significantly affect the system's behavior when the field is swept antiparallel to the skyrmion-core polarization, they compromise the stability of the SkL phase and dramatically modify the associated magnetization processes when the field is swept parallel. In fact, the annihilation of the lattice starts at the defect site. At high PMA, where the inversion does not occur, an inhomogeneous skyrmion growth with twisting and deformation of skyrmions is promoted around the defect (Figure 5a-c), resulting in elliptical instabilities \cite{Moutafis2009} and consequently a gradual loss of all topological charge (Figure 5d). At low PMA, where inversion occurs in ideal systems, lattice destruction and charge leakage start around the defect (Figure 5e-f), but not all topological charge is destroyed because some skyrmions survive the inversion (Figure 5g-h), suggesting that inversion could be experimentally observed even in defective systems. 

Importantly, the critical field in which the lattice is destroyed is strongly reduced (see Figure 5i), and the transition changes from first-order to second-order, exhibiting lattice-melting behavior. This destabilization of skyrmion lattices due to the presence of defects illustrates the importance of symmetry in the system. In an ideal SkL, the application of a parallel field leads to the hexagonal state, where skyrmions are stabilized by a shared 2$\pi$-domain-wall network, i.e., the skyrmion lattice is protected by the topology of both the skyrmions themselves and the 2$\pi$ boundaries. This suggests that the topological charge of an ideal lattice is \emph{delocalized} when the constraint is enforced by the 2$\pi$-domain-wall network. In contrast, in a system with even a single defect, the lattice is destabilized because the topological protection provided by the boundaries of the skyrmion at the defect site is lost. The defect site therefore acts as a topological-charge sink, and thus the stability of skyrmion lattices crucially depends on the density of defects.

\section{Conclusions}

Our study reveals the complex underlying mechanisms of topological-charge creation and annihilation in thin magnetic films with the Dzyaloshinskii-Moriya interaction, both isotropic (bulk) and anisotropic (interfacial), and perpendicular magnetic anisotropy. We have found that upon the application of an external field in ideal infinite films the skyrmion lattice phase undergoes a first-order phase transition, either to the topologically trivial ferromagnetic state or to an inverted skyrmion lattice phase via the transient formation of antiskyrmions. The first-order character of any phase transition in an ideal lattice is due to the delocalization of the topological charge within a 2$\pi$-domain-wall network and the consequent collective response to external fields. In the presence of even a single defect, however, the skyrmion lattice phase is unstable and collapses gradually through a defect-induced melting process, where the defect site acts as a topological-charge sink. These findings emphasize the importance of imperfections in materials and their implications on the stability of topologically non-trivial spin textures, demonstrating that the consideration of defects is paramount for the analysis of experimental data. This provides a basis for a much wider scope of experiments on skyrmion lattices, particularly concerning the development of materials for skyrmion-based devices.


\section*{Acknowledgments}
LP, MC and JFL gratefully acknowledge funding from the Swiss National Science Foundation (Grant No. 200021--172934). MC and CM thank the Royal Society International Exchanges programme (Ref: IE161506).

\section*{Supplementary Information}

\begin{figure}[h]
\centering
\includegraphics[width=1\columnwidth]{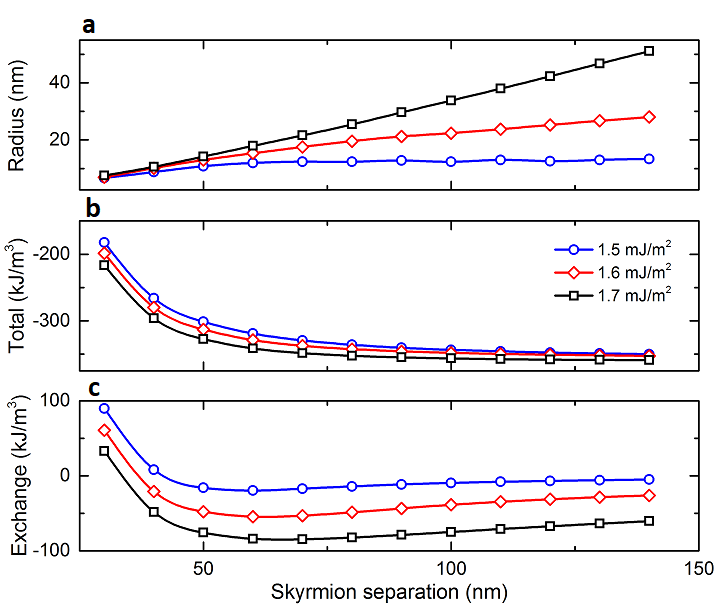}
\caption{Plots of \textbf{a} skyrmion radius, \textbf{b} total energy density and \textbf{c} exchange energy density against the skyrmion separation in the lattice for three different values of DMI strength. At strong DMI and low PMA (the latter result not shown in the graph) the skyrmion radius grows with the skyrmion separation. This might be promoted by the existence of an exchange energy minimum, whose position is around 50 - 60 nm and increases for stronger DMI (black and red lines) and lower PMA. The growth is inhibited at low DMI and high PMA, probably because the minimum in exchange energy vanishes (blue line).}
\label{DMI_density}
\end{figure}

We have investigated the effect of the separation between skyrmions on their size, and we found that at low PMA and strong DMI they grow linearly with the separation. The energy density analysis shows that there is a minimum in the exchange energy at the separation of 50-60 nm, which is associated to the growth (Fig \ref{DMI_density}). The separation at which the minimum exists is larger in materials with lower PMA and stronger DMI. Contrarily, the minimum does not exist at at high PMA and weak DMI, where the growth of skyrmions is inhibited after they reach the radius of about 15 nm. 

\end{document}